\documentclass[aps,prl,showpacs,twocolumn,groupedaddress,amssymb]{revtex4}

\usepackage{graphicx}
\usepackage{dcolumn}
\usepackage{bm}

\begin{document}

\title{THz radiation
using the two plasmon decay and the backward Raman scattering}

\author{S. Son}
\affiliation{18 Caleb Lane, Princeton, NJ, 08540}

\begin{abstract}
A scheme of  THz radiation using two moderately intense lasers and a moderately relativistic electron beam is 
proposed.  In the scheme, 
a laser encounters a co-propagating relativistic electron beam,
and excites plasmons via the two-plasmon decay. 
The excited plasmons will emit the THz radiations, interacting with the second laser via the Raman scattering. 
 Our estimation suggests that the mean-free path   of the pump laser
 to the THz radiation is  much shorter than the Thomson scattering. 
The physical parameters for practical interests are presented. 
\end{abstract}

\pacs{42.55.Vc, 42.65.Ky, 52.38.-r, 52.35.Hr}       

\maketitle

An intense THz light source with frequency between 1 and 10 THz has many
applications~\cite{radar,diagnostic,seigel3,security}.
However, significant progresses over the intensity of the THz light need to be made before those applications are realized, because there are various limitations of the available technologies~\cite{siegel, siegel2, booske, gyrotron, gyrotron2, gyrotron3, magnetron, qlaser, qlaser2, freelaser, freelaser2}; 
The current inability to produce high intense THz light comparable to the theoretical limit is referred to as the ``THz gap''~\cite{siegel, siegel2, booske}.

One emerging way to generate intense THz light is the scattering between an  electron beam and an intense laser~\cite{sonttera, sonltera}, potentially surpassing the conventional free electron laser (FEL)~\cite{freelaser, freelaser2}. 
Those schemes become plausible thanks to the great advance in the electron beam~\cite{monoelectron, ebeam} and the intense laser~\cite{cpa, cpa2, cpa3,cpa4}, which are being
developed 
for the inertial confinement fusion~\cite{tabak,sonprl,sonpla}.  
In this paper, 
we propose a new scheme of  intense THz radiation 
based on the laser-plasma interaction between an electron beam and a laser.

Consider a situation when an intense laser (plasmon pump) 
encounters a moderately relativistic electron beam moving in the \textit{same direction}. 
Denote the electron density and velocity (the relativistic factor) 
of the beam as $n_0$ and $v_0$ ($\gamma_0^{-1}= \sqrt{1 - \beta^2}$, $\beta = v_0 /c $).  
If the laser frequency satisfies the following condition: 
\begin{equation} 
\omega_{pp0} \cong (2 + \sqrt{3} \beta) \sqrt{\gamma_0} \omega_{pe} 
\mathrm{,} \label{eq:pump} 
\end{equation} 
where $\omega_{pp0}$ is the laser frequency and 
     $\omega_{pe}^2 = 4\pi n_0 e^2 /m_e$ is the plasmon frequency, 
the laser  excites  the plasmons 
via the well-known two plasmon decay~\cite{liu, langdon}. 
Then, the second laser (BRS pump laser), injected into the electron beam in the \textit{same direction}, 
 could emit the THz radiations via the BRS 
between the plasmon pump laser and the plasmons excited. 
The THz radiation (the seed pulse) will be emitted  
in the \textit{opposite direction} to the electron  beam. 
 Due to the relativistic velocity of the electron beam, 
the ratio of the seed pulse frequency to the BRS pump laser frequency  is given as   
\begin{equation} 
\frac{\omega_{s0}}{\omega_{p0}}  \cong   \frac{ 
\sqrt{(1 + 3 P^2)} - \sqrt{3} \beta P}{ 
{\sqrt{1 + 3 S^{2}} +\sqrt{3} \beta S }} \label{eq:down2} \mathrm{,}
\end{equation}
where $P= S - 1/\sqrt{3} $. $\omega_{s0}$ ($\omega_{p0}$)  is the 
frequency of the seed pulse (pump pulse) and $S$ is defined as
\begin{equation} 
 S =  \left(\frac{\omega_{p0}}{\gamma_0^{1/2}  \omega_{pe}} 
  - \frac{\gamma_0^{1/2}  \omega_{pe}}{\omega_{p0}} \right)/ 2 \sqrt{3}
\mathrm{.}\label{eq:S}
 \nonumber
\end{equation}
where it is assume that $\beta \cong 1$.
In this paper, we derive 
Eqs.~(\ref{eq:pump})  and (\ref{eq:down2}) and 
analyze the practical plausibility of the current scheme. 
We suggest that the optimal value of $S$ should be $1.0 < S < 10.0 $, and the electron beam density  (the beam relativistic factor) should be of $10^{15} / \mathrm{cc} $ to 
$10^{17} / \mathrm{cc} $ (1 to 10). 
The required intensity of lasers is shown to be considerably lower 
than  other  schemes of the THz radiation, 
The conversion efficiency of the pump laser energy to the THz radiation 
is shown to be a few pecents, which is ultra higher than other schemes. 
The disadvantages or advantages over various other schemes are discussed.   

To begin with,
consider a rather uniform relativistic electron beam with the density $n_0$ 
in the laboratory frame and 
the relativistic factor $\gamma_0$.  
The plasmon pump laser propagates in the same direction with the electron beam. 
It is most convenient to 
describe the physics in the co-moving reference frame 
where the electron beam is stationary.  
From now on, we denote the laboratory frame (the co-moving frame) as 0 (1); 
For an example, the wave frequency and the wave vector in the laboratory frame (the co-moving frame) are denoted as 
 $\omega_{pp0}$  and $k_{pp0} $ ($\omega_{pp1}$  and $k_{pp1} $). 
In the co-moving frame, the electron density is given as 
$n_1 = n_0 / \gamma_0 $ due to the length dilation and  
a photon satisfies the dispersion relation ship, 
$\omega^2 =  \omega_{pe}^2/\gamma_0 + c^2 k^2 $. 
The two plasmon decay occurs when $\omega \cong 2\omega_{pe}/\sqrt{\gamma_0}$ in the co-moving frame or $ ck \cong \sqrt{3} \omega_{pe} / \sqrt{\gamma_0}$. 
Using the photon dispersion relationship and the Lorentz transform, 
the wave vectors (wave frequencies) between the laboratory frame and the co-moving frame are related as

\begin{eqnarray} 
\omega_{pp0} &=& \gamma_0 \left[ \sqrt{\omega_{pe}^2/\gamma_0 + c^2 k_{pp1}^2 } + vk_{pp1} \right] \mathrm{,}  \label{eq:lorentz1} \\  \nonumber \\
k_{pp0} &=&  \gamma_0 \left[ k_{pp1} + \frac{\omega_{pp1} }{c}  \frac{v_0}{c} \right] \mathrm{,} \label{eq:lorentz2}
\end{eqnarray}
where $\omega_{pp0}$ and $k_{pp0}$  ($\omega_{pp1}$ and $k_{pp1}$)  
are the wave frequency and the vector of the laser
 in the laboratory frame (the co-moving frame).  
From Eq.~(\ref{eq:lorentz1}) and  
the condition of the two-plasmon decay 
($ck_{pp1}  \cong \sqrt{3} \omega_{pe} / \sqrt{\gamma_0}$), 
we derive  Eq.~(\ref{eq:pump}).
In the two-plasmon decay, the plasmons with the angle $\pi/4$ 
to the laser direction  are excited most strongly~\cite{liu}. 
Our primary interest is the plasmon with the wave vector in the parallel
 direction  to the electron beam. 
For the strong plasmon in that direction, 
the laser should  be in the $\pi/4$ angle 
to the electron beam direction in the co-moving frame~\cite{liu} or  
 $k_{pp1x} / k_{pp1z} \cong 1 $ where $k_{pp1x} $ 
and $k_{pp1z} $ is the wave vector in the x and z direction 
(the beam direction is assumed to be the z-direction). 
The $k_{pp0z} $ can be obtained from Eq.~(\ref{eq:lorentz2}) while $k_{pp1x} = k_{pp0x} $, and then 
 $ck_{pp0x} \cong \sqrt{1.5} \omega_{pe} / \sqrt{\gamma_0}$ 
and $ck_{pp0z} \cong \gamma_0 ( \sqrt{1.5} + 2 \beta)/  \sqrt{\gamma_0}$ 
so that  $k_{pp0z} / k_{pp0x} \cong \gamma_0 ( 1 + \sqrt{8/3} \beta) $. 

As the plasmon pump  laser satisfies 
 $\omega_{pp1} \cong 2 \omega_{pe} / \sqrt{\gamma_0} $ in the co-moving frame,  
it will excite the plasmons via the two-plasmon decay. 
The density fluctuation due to laser  is well-analyzed and given as~\cite{liu} 

\begin{eqnarray} 
\left(\frac{\delta n_1}{n_1 }\right)^2 &\cong& 
  \frac{3}{8 \pi} \left( \frac{c}{v_{te} } \right) \left(\frac{c^2k_{pp1}^2 \gamma_0}{\omega_{pe}^2}\right)
   \left(\frac{e^2E_{pp1}^2 \gamma_0}{m_e^2\omega_{pe}^2 c^2}\right) \nonumber \\ \nonumber \\
 &=& \frac{9}{2\pi} \gamma_0^2\left( \frac{c}{v_{te} } \right)  
\left(\frac{v_q^2 }{c^2}\right)\left( \frac{k_{3}^2}{k_{pp1}^2}\right) \mathrm{,}  \label{eq:lan} 
  \end{eqnarray}
where $E_{pp1}$ is the electric field strength of the plasmon pump laser in the co-moving frame, and 
$v_q = e E_{pp1} / m_e \omega_{ppl}$ is the quiver velocity,  $k_3$ is the wave vector of the plasmon, 
$v_{te} $ is the electron thermal velocity in the same frame, and we use 
$ck_{pp1} \cong \sqrt{3} \omega_{pe} / \sqrt{\gamma_0}$ and $\omega_{pp1}= 2ck_{pp1} / \sqrt{3} $.
The threshold condition for the two-plasmon decay is given as $1/3 (v_q/v_{te})^2 k_{pp1} L > 1$, where
$L$ is the length scale of the density variation.  
For an example,  for the co2 laser with $k_{pp1} L \cong 100 $ and 1 keV electron plasma, 
the threshold intensity is given as $I \cong 10^{10}  \ \mathrm{W} / \mathrm{cm}^2 $ when $k_{3}/k_{pp1} \cong 3$. 
One useful fact is that the quiver velocity $v_q $
 is invariant under the Lorentz transform. 
Also note that the kinetic energy spread $\delta E / E $ of the electron beam 
in the laboratory frame is the same order with the velocity spread of the beam in the co-moving frame: 
 $\delta E/ E \cong \delta v / v$. 
Assuming  the beam energy spread in the laboratory frame is 
between 1 \% and 10 \%, the electron temperature 
in the co-moving frame is between 50 eV and 5 keV.

If  plasmons with considerable intensity are 
excited by the plasmon pump laser as in Eq.~(\ref{eq:lan}), and
 another laser (BRS pump laser)  is injected into the beam in the same direction with the electron beam,  
the RBS pump laser and the plasmons could induce 
the THz radiation via the BRS. 
The energy and momentum conservation of the BRS is given as 
\begin{eqnarray} 
 \omega_{p1} &=& \omega_{s1} + \omega_{3} 
\mathrm{,} \nonumber \\
 k_{p1} &=& k_{s1} +k_3 \mathrm{,} \label{eq:cons}  
\end{eqnarray}
where  $\omega_{p1} $ and  $k_{p1}$ ($\omega_{s1} $ and  $k_{s1}$)  is the wave frequency and vector of the BRS pump laser (the seed pulse) in the co-moving frame, 
and $k_3$ and $\omega_{3} $ is the wave vector and the wave frequency 
of the Langmuir wave excited: 
$\omega_3 \cong \omega_{pe} / \sqrt{\gamma_0} $.  
It is usually the case that $k_{p1} > k_{pp1}$ and thus  
define $ S = k_{p1} / k_{pp1} > 1$. Then 
the wave frequencies  of the pump pulse (the seed pulse) between the co-moving frame and the laboratory frame are related from the Lorentz transform: 
\begin{eqnarray}
\frac{\omega_{s0}}{\sqrt{\gamma_0} \omega_{pe} } 
 &=& \left( \sqrt{1 +P^2 } - \sqrt{3} \beta P \right)  = \Delta_p  \mathrm{,}\nonumber \\  \nonumber \\ 
\frac{\omega_{p0}}{\sqrt{\gamma_0} \omega_{pe} }
 &=& \left( \sqrt{1 + 3 S^2 } + \sqrt{3} \beta S  \right)  = \Delta_s \label{eq:seed} \mathrm{,} \\ \nonumber 
\end{eqnarray}
where we use $ P = k_{s1} / k_{pp1} \cong (k_{p1}-\omega_{pe} 
/\sqrt{\gamma_0}) / k_{pp1}) \cong  S - 1/\sqrt{3} $ (valid when $S - 1/\sqrt{3} >  1$).
From Eq.~(\ref{eq:seed}), Eq.~(\ref{eq:down2}) and  Eq.~(\ref{eq:S}) can be derived.

The THz radiation induced by the BRS from the pump pulse and the plasmon 
is described \textit{in the co-moving frame} as follows~\cite{McKinstrie}.
\begin{equation}
\left( \frac{\partial }{\partial t} + v_s \frac{\partial}{\partial x} + \nu_2\right)A_s  = -ic_s A_p A^*_3   \label{eq:2} \mathrm{,}
\end{equation}
where $A_i= eE_{i1}/m_e\omega_{i1}c$  is 
the ratio of  the electron quiver velocity of the pump pulse ($i=p$)
and seed pulse ($i=s$), relative to the velocity of the light $c$, 
 $A_3 = \delta n_1/n_1$ is the the Langmuir wave amplitude,
$\nu_2$ is the rate of the inverse bremsstrahlung  
of the seed and 
$ c_2 = \omega_3^2/ 2 \omega_{p1}$.
From Eq.~(\ref{eq:2}),  
the considerable part of the pump energy will be transferred to the 
seed pulse when $c_s A_3 \delta t_b \cong 1$: 
\begin{equation} 
l_b = \delta t c  \cong c (2 \omega_{s1}  /\omega_3^2 ) (1/A_3) 
\mathrm{.} \label{eq:mean}
\end{equation}
 On the other hand, 
the Thomson scattering suggests that  $ l_t \cong 1/n\sigma_t $ with $\sigma_t = (mc^2 / e^2)^2 $.  For an example, when $n_1 \cong 10^{16} / \mathrm{cc} $,   
we estimate $l_t \cong 10^9 \ \mathrm{cm} $ and $l_b \cong (10^{-2} / A_3) S^2 \ \mathrm{cm} $.  Even for $A_3 \cong 0.001 $, the THz radiation by the BRS is considerably stronger than the Thomson scattering or $l_t \gg l_b $

The maximum conversion efficiency from the pump energy to the seed energy can be estimated as follows. 
Denote the total energy of the BRS pump laser (the seed laser) \textit{in the laboratory frame}
 as $\mathrm{E}_{p0} $ ($\mathrm{E}_{s0}$).  
\textit{In the co-moving frame}, 
the BRS pump energy is seen to be  
$\mathrm{E}_{p1}  \cong (\sqrt{3} S \mathrm{E}_{p0}/ 
\Delta_s \gamma_0 )$ 
from Eq.~(\ref{eq:seed}). 
Considering the conversion efficiency in this co-moving as $\epsilon_1 $, 
the energy of the seed pulse is given as 
  $\mathrm{E}_{s1} =  \epsilon_1 (\Delta_p\gamma_0 \mathrm{E}_{p1}/\sqrt{3}S ) $. 
This energy of the seed pulse is seen \textit{in the laboratory} to be 
   $\mathrm{E}_{s0} = \epsilon_1
\Delta_p \gamma_0 \mathrm{E}_{s1}  /\sqrt{3}S\cong \epsilon_1 (\Delta_p / \Delta_s) \mathrm{E}_{p0}$. 
Then, the conversion efficiency in the laboratory frame is given as 

\begin{equation} 
\epsilon_0 =    \left(\frac{\Delta_p }{\Delta_s} \right) \epsilon_1  
\cong  \left( \frac{ 
\sqrt{(1 + 3 P^2)} - \sqrt{3} \beta P }{ 
{\sqrt{(1 + 3 S^{2})} +\sqrt{3} \beta S }}\right) \epsilon_1  \mathrm{.}
\label{eq:conv} 
\end{equation}
The estimation of  $\epsilon_1$ in the co-moving frame follows. 
If the plasmons excited by the two-plasmon decay are isotropically distributed, 
the radiation by the BRS would be isotropic as  in the Thomson scattering. 
However, only the photon in the direct opposite direction to the beam direction would be down-shifted to the THz range in the laboratory frame; 
The angular width, that are relevant to the THz, will be  
 $d \theta \cong  S/ \Delta_s \gamma_0 $ in the co-moving frame.  
Then, the conversion efficiency $\epsilon_1$  would be   $\epsilon_1 
\cong ( S/ \Delta_p \gamma_0 )^2$. 
On the other hand,  if the angular distribution of the plasmons are sharply peaked at $\theta = 0 $, 
the most of the pump pulse will be radiated into the opposite direction to the beam, in which case  $\epsilon_1 \cong 1 $.  
From the above consideration, we could estimate the optimal  conversion efficiency as 

\begin{equation}
 \left(\frac{S^2}{\Delta^2_p \gamma_0^2 }\right)\left(\frac{\Delta_p}{ \Delta_s }\right) <  \epsilon_0  <  \frac{\Delta_p}{ \Delta_s } \mathrm{.}  \label{eq:eps}
\end{equation}
Eq.~(\ref{eq:eps}) is the maximum possible efficiency since we assume that most of the BRS pump energy will be radiated via the BRS scattering. 
In Eq.~(\ref{eq:eps}), it is assumed that $d \theta = S/ \Delta_s \gamma_0 < 1$. If  $S/ \Delta_s \gamma_0>1$ as is often the case,  then 
$\epsilon_0 \cong  \Delta_p / \Delta_s $.

We provide the estimation of the gain, the mean-free path and the THz frequency in  a few  examples of  the electron beams and the lasers
 in Table~(\ref{tb}). 
From the estimations, 
we conclude that, for the most effective THz radiation, 
the $S$ ($\gamma_0$) should be in the range $1<S<5 $ ($2<\gamma_0<10$) and that 
the plasmon (BRS) pump laser could be the co2 or ND:YAG laser but preferentially the co2 laser. 
 The THz photon generated ranges from $0.5 \ \mathrm{THz} $ to  $10  \ \mathrm{THz}$.
The conversion efficiency could be as high as a few percents 
and the BRS wave length is always shorter than the Thomson scattering in those examples.

\begin{table}[t]
\centering
\begin{tabular}{|c||cccc||}
	\hline
   &  1    &   2  & 3 & 4    \\
	\hline \hline 
  $n_{16}$       & 16 & 5.26 & 46 & 13.72 \\ 
 $\gamma_0$     & 5 & 15 & 2 & 100  \\ 
 $S$            & 4.42   & 9.42 & 2.42 &  3 \\ 
$\lambda_{pp0}$ & 10.9 & 10 & 9.95 & 2.41  \\
$\lambda_{p0} $  & 2.45 & 1.15 & 7.27 & 1.1 \\ 
$\mathrm{F_{s0}}$    & \textbf{1.87} & \textbf{0.53} & \textbf{5.21} & \textbf{3.17}
  \\
$A_3$   & 0.2 & 0.45& 0.067 & 0.027
  \\ 
$l_t$  &9.15  & 85 & 1.3 & 218
  \\ 
$l_b$ &  0.09 & 0.27 & 0.034 & 1.81 \\ 
$\Delta_p / \Delta_s$ & 65 & 484 & 7.9 & 89 \\
$\sqrt{S} / \Delta_p \gamma_0$ & 3.7 & 9.2 & 1.15 & 0.24 \\
\hline
\end{tabular}
\caption{The laser and electron beam parameter and the 
characteristic of the THz radiation \label{tb}
In the table, $n_{16} $ is the electron density $n_0$ normalized by $10^{16} / 
\mathrm{cc} $,  $S $ is defined in Eq.~(\ref{eq:S}), $\lambda_{pp1} $ and $\lambda_{p0} $ is the wave length of  the plasmon and BRS pump laser normalized by THz, 
 $F_{s0} = \omega_{s0} / 2 \pi $ is 
the frequency of the seed pulse normalized by $10^{12} / \sec$
from Eq.~(\ref{eq:down2}),  $A_3=\delta n_1 / n_1$ is the plasmon intensity given in Eq.~(\ref{eq:lan}), $l_t$ ($l_b$) is the mean-free path of the Thomson scattering  (BRS) in the unit of cm 
 obtained from Eq.~(\ref{eq:mean}), 
$\Delta_p / \Delta_s $ is the ratio of the seed pulse frequency to the BRS pump pulse frequency as given in Eq.~(\ref{eq:down2}). 
In this example, we consider the plasmon pump intensity of $I= 10^{11} \  \mathrm{W} / \sec$ and the electron temperature of the beam in the co-moving frame is assumed to be 5 KeV. 
}
\end{table}

In summary, we propose a scheme of THz radiations. 
The scheme is based on the two-plasmon decay and the backward Raman scattering. 
The first laser excites plasmons via the two-plasmon decay in a moderately relativistic electron beam, 
and the second laser excites the BRS and emits the THz radiation 
in the opposite direction to the beam. 
The estimation suggests that 
the laser wave length between 1 $\mu$m and 10 $\mu$m can be used as 
the BRS and plasmon pump lasers. 
 and that the optimal electron beam has 
the  density of $10^{15} / \mathrm{cc} $ to 
$10^{17} / \mathrm{cc} $ and the relativistic factor of 2 to 10.
The frequency of the THz radiation would be between 0.5 THz and 10 THz. 
We estimate the conversion efficiency as high as  a few \%. 
The mean-free path of the BRS is shown to be much shorter than the Thomson scattering. 

In comparison with other THz schemes using the laser-plasma interaction~\cite{sonttera, sonltera}, 
the threshold intensity (conversion efficiency) of the pump laser is  lower (higher) than the other schemes. 
However, the scheme proposed has a few drawbacks. 
First, an uniform electron  beam is needed for strong two-plasmon decay. 
Second, the beam size should be rather long for high conversion efficiency. 
Third, once the electron beam characteristics 
such as the density and the relativistic factor are fixed, 
the plasma pump frequency cannot be adjusted as given in Eq.~(\ref{eq:pump}). 
However, even with all these drawbacks, 
the scheme could be very attractive as a THz source due to the lower laser intensity threshold and the possible high conversion efficiency.

\bibliography{tera2}

\end{document}